\documentclass[12pt]{article}

\usepackage{amsmath}
\usepackage{amssymb}
\usepackage{graphicx}
\usepackage{color}
\usepackage{mathabx}
\usepackage{hyperref}

\usepackage{xr}
\numberwithin{equation}{section}
\newcommand{\tr}{\mbox{Tr}}

\newcounter{resultcounter}[section]

\newtheorem{thm}[resultcounter]{Theorem}

\newcommand{\h}{{\cal H}}

\newcommand{\rx}{{\mathbb R}}

\def\qed{\hfill $\Box$\medskip}
\newcommand{\bbbone}{\mathchoice {\rm 1\mskip-4mu l} {\rm 1\mskip-4mu l}
{\rm 1\mskip-4.5mu l} {\rm 1\mskip-5mu l}}

\begin{document}

\title{Stability of PPT in equilibrium states
	\bigskip}

\author{ Marco Merkli\footnote{merkli@mun.ca} \ and \ Mitch Zagrodnik\footnote{mzagrodnik@mun.ca}\bigskip\\
Department of Mathematics and Statistics\\
Memorial University of Newfoundland\\ St. John's, NL, Canada A1C 5S7}  
\maketitle
\vspace*{-.3cm}

\medskip
\begin{abstract}
We use simple spectral perturbation theory to show that the positive partial transpose property is stable under bounded perturbations of the Hamiltonian, for equilibrium states in infinite dimensions. The result holds provided the temperature is high enough, or equivalently, provided the perturbation is small enough. 
\end{abstract}

\section{Introduction}

The positive partial transposition (PPT), or Peres-Horodecki criterion, gives a necessary condition for a density matrix $\rho$ of a bipartite quantum system SB (`system-bath') to be separable. Namely, if $\rho$ is separable, then the partial transpose (relative to either subsystem, say B), $T_{\rm B}[\rho]$ is a positive operator \cite{Peres,HHH96}. Equivalently, if $\rho$ is not PPT, then it must be inseparable, also called entangled. If the dimension of the subsystem Hilbert spaces are two or three, then the converse statement is also correct \cite{HHH96}. However, in dimension $\ge 4$, there are states which are PPT and yet entangled. In this situation the entanglement is called  bound. It is called so since one cannot extract from such a state, pure singlet (entangled) states, which would be needed as a resource for quantum information purposes. In contrast, if it is possible to extract (by LOCC protocols) from an entangled mixed state (or many copies thereof), pairs of particles in a pure singlet state, then the mixed state is said to have  `distillable' entanglement  \cite{Bennet96}. The corresponding state is said to be distillable. It is shown in \cite{HHH98} (for finite dimensional systems) that if a state is PPT then it is not distillable. The converse is not true in dimension $3$ or higher, though. Any possible entanglement in a PPT state is  bound. For qubits (dimension $2$) it was shown in \cite{HHH97-2} that any entangled state is distillable. However, in higher dimensions there are entangled states with bound entanglement, as shown in  \cite{WW}.
\medskip

A great feature of the PPT criterion is that in principle, it is easy to apply. To verify that a given density matrix $\rho$ of a bipartite system is entangled, one `simply' has to check that the partial transpose $T_B[\rho]$ is not a positive operator. We are addressing the following question here:
\medskip

 {\em Suppose a density matrix $\rho_0$ is PPT, and consider a modified density matrix $\rho=\rho_0+\rho'$, where $\rho'$ is a perturbation operator. Under what conditions is $\rho$ still PPT?}
\medskip

To investigate the question we first note that since $\rho$ is hermitian ($=$ selfadjoint) then so is $T_B[\rho]$ (see also \eqref{2.14-1} below). Thus  $T_B[\rho]\ge0$ if and only if all the eigenvalues of $T_B[\rho]$ are non-negative. The partial trace is a linear operation,  $T_B[\rho]=T_B[\rho_0]+T_B[\rho']$. Basic perturbation theory \cite{Kato} tells us that the eigenvalues of $T_B[\rho]$ lie within a neighbourhood of the size $\|T_B[\rho']\|_\infty$ (operator norm of the perturbation) of the eigenvalues of $T_B[\rho_0]$.  Since $\rho_0$ is PPT we know that  $T_B[\rho_0]\ge 0$. If $T_B[\rho_0]$ has a lowest eigenvalue $\lambda_0>0$ then for $\|T_B[\rho']\|_\infty<\lambda_0$ the spectrum of $T_B[\rho]$ is guaranteed to be $\ge 0$, which means that $\rho$ is PPT. 

This straightforward approach breaks down as soon as the dimension is infinite. The reason is that $\rho_0$ and hence $T_B[\rho_0]$ are Hilbert-Schmidt operators ({\em c.f.} \eqref{2.14}). In particular, $T_B[\rho_0]$ is a compact operator and therefore, in the infinite-dimensional case,  its eigenvalues must accumulate at the origin. Consequently, no matter how small we take $\|T_B[\rho']\|_\infty$, the simple argument given above does not work. 
\medskip

We show in this paper how one can modify the simple perturbation argument for equilibrium states, where the perturbation is a bounded interaction term $V$ in the Hamiltonian,
\begin{equation}
	\label{setup}
\rho_0 = \frac{e^{-\beta H_0}}{\tr e^{-\beta H_0}},\qquad \rho= \frac{e^{-\beta (H_0+V)}}{\tr e^{-\beta(H_0+V)}}
\end{equation}
with 
$$
H_0= H_A\otimes\bbbone_B + \bbbone_A\otimes H_B.
$$
Thermal states form the cornerstone of equilibrium statistical mechanics. Examining their quantum properties, one of which is measured by entanglement, is an important question. That said, the equilibrium setup \eqref{setup} is mathematically quite general. Indeed, any faithful  density matrix\footnote{A density matrix is called faithful if zero is not one of its eigenvalues.} $\rho_0$  can be written in the form $\rho_0 = e^{-H_0}/{\tr e^{-H_0}}$ for some Hermitian (self-adjoint) $H_0$, and similarly for $\rho$. Our approach is  applicable to any such $\rho_0, \rho$.
In this work, we express the perturbation $\rho_0\mapsto \rho$ as an interaction term $V$ in the Hamiltonian because this is physically intuitive. Our main idea is to use the Dyson expansion to write
\begin{equation}
    \label{x1}
e^{-\beta (H_0+V)} = e^{-\beta H_0/2} \big[\bbbone +\mathcal O(V)\big]e^{-\beta H_0/2},
\end{equation}
where $\mathcal O(V)$ is an operator which vanishes for $V=0$. PPT for  $\rho$ will then follow provided $T_B[\mathcal O(V)]$ is small enough such that 
$$
\bbbone+T_B[\mathcal O(V)]\ge 0.
$$
By factoring out the operators $e^{-\beta H_0/2}$ in \eqref{x1} we remove the problem of eigenvalues accumulating at the origin, as the unperturbed density matrix is effectively replaced by the operator $\bbbone$ now, whose spectrum $\{1\}$ is separated from the origin. The detailed control of $\mathcal O(V)$ leading to the wanted bounds involves the size of $V$ as well as the inverse temperature $\beta$. We assume a bound on the Hilbert-Schmidt norm of the imaginary time evolved interaction operator. Namely, we assume that there are constants $a,b$, and $s_*>0$ such that for all $0\le s\le s_*$, we have the bound
$$
	\| e^{-s H_0}Ve^{sH_0} \|_2 \le a e^{bs}.
$$
We then show in Theorem \ref{thm1} that for large enough temperature $\beta\le \max\{\beta_*, s_*\}$, the perturbed state $\rho$ is PPT. The upper bound $\beta_*$ on $\beta$ depends on the constants $a,b$. If $a$ is small (say $V$ contains an overall small coupling constant) then $\beta_*\sim \ln(1/a)$. So the smaller $a$ is the larger we can take $\beta$ (the smaller we can take the actual temperature $1/\beta$) for the result to hold. Conversely, if the coupling $V$ is not small ($a$ sizeable) then the temperature has to be higher for the validity of our derivation.

\bigskip
%\color{cyan}
{\bf Literature.} The question of separability of thermal equilibrium states has a rich history. Entanglement is a measure for the degree of their ``quantumness''. Intuitively it is expected that a quantum to classical transition happens at high temperature $T$, that is, entanglement disappears for large $T$. From a quantum information point of view, this means that only cool enough materials may be used as a  quantum resource.

Many works deal with spin (qubit) chains in equilibrium. They analyze how 2-qubit entanglement along the chain depends on various parameters \cite{Arnesen2001,Gunlycke2001, X2001, X2001-2, X2002, Osborne2002, X2002-2, Glaser2003, Brennen2004,Panate}, in particular finding temperature bounds which guarantee entanglement or separability. In this situation, one can conveniently use concurrence to quanitfy entanglement. An analysis for two spins of arbitrary length (where concurrence cannot be used as an entanglement measure) was carried out in \cite{Schliemann}. In that work, the validity of the PPT criterion was linked to properties of the spin correlators.  More generally, it was shown in \cite{Raggio2005, Raggio2006} that for finite dimensional systems, thermal states with respect to {\em any} Hamiltonian are separable at high enough temperature, $T\ge T_c$, for some critical $T_c$ and that any interval $I\subset (0,T_c)$ contains a $T'$ such that the equilibrium state at temperature $T'$ is entangled. A general topological argument,  which again works for finite-dimensional systems, was presented in \cite{FMB}: The equilibrium density matrix at infinite temperature $T\rightarrow\infty$ is proportional to the identity matrix. The latter is a product state (under any bi-partition of the total system) and it is contained in a ``ball'' (topological neighbourhood) of separable states. As the equilibrium density matrix depends continuously on the inverse temperature $\beta=1/T$, the infinite temperature density matrix cannot be transformed into an entangled density matrix by an infinitesimal change of $\beta$ away from $\beta=0$. Hence there must be a critical temperature $T_c$ so that for $T>T_c$ the state is separable.

In all the works above, entanglement between finite-dimensional systems is studied. The advantage of our method is that it is very simple and works in infinite dimensions. However, we only show that the PPT property holds at high enough temperatures, and this gives only partial information on the entanglement. Namely, we do not settle the question of bound entanglement; our result presented here does not show whether there exists entanglement in the regime where PPT is satisfied. In this sense, our results are more modest than many of the ones cited above. The presence of bound entanglement in thermal spin states (finite dimensions) was derived in \cite{Toth,Cavalcanti2010}. The authors of \cite{Audenaert} consider the thermal states of a closed chain of harmonic oscillators and find an explicit expression for the logartihmic negativity. This allows them to discuss entanglement properties for this specific infinite-dimensional system. Based on this work, the existence of bound entanglement in the same system was shown in \cite{Cavalcanti2008}. A more general approach exhibiting bound entanglement for infinite-dimensional systems would be valuable. 

Finally, it is interesting to note that there are models where entanglement survives `at all temperatures'. This was shown to hold for a mirror in thermal equilibrium interacting with an electromagnetic field mode in a coherent state \cite{Ferreira}. 

%\color{black}

\color{cyan}

\color{black}

\section{PPT criterion \& main result}

We start out by defining the notions involved in the PPT, or {\em positive partial transpose} criterion and state that criterion below in Theorem \ref{thm1}. 

Let $\h$ be a separable Hilbert space, $\dim\h\le \infty$. Our main focus will be on the infinite dimensional case. The norm of a vector $|\psi\rangle\in \h$ is given by $\|\, |\psi\rangle\, \| = \sqrt{\langle \psi|\psi\rangle}$, where $\langle\cdot |\cdot\rangle$ is the inner product of $\h$. Let $\mathcal B(\h)$ denote the set of all bounded linear operators on $\h$. We use the following  three norms of operators $X\in\mathcal B(\h)$,
$$
\|X\|_\infty=\sup_{|\psi\rangle\in\h,\  \| |\psi\rangle\|=1} \|X|\psi\rangle\|,\qquad \|X\|_2=\big(\tr\, |X|^2\big)^{1/2}, \qquad \|X\|_1=\tr|X|,
$$
where $|X|=\sqrt{X^\dag X}$ and $X^\dag$ denotes the adjoint of $X$. The norms are also called the operator norm ($\|X\|_\infty$), the Hilbert-Schmidt norm ($\|X\|_2)$ and the trace norm ($\|X\|_1$). The inequalities $\|X\|_\infty\le \|X\|_j$, $j=1,2$ are well known. We denote by $\mathcal T_2(\h)$ all bounded operators $X$ such that $\|X\|_2<\infty$. These are called the Hilbert-Schmidt operators.  $\mathcal T_2(\h)$ is Hilbert space when equipped with the inner product $\langle X,Y\rangle = \tr\, (X^\dag Y)$. The collection of all trace class operators in $\h$ (operators with finite trace norm) is denoted by $\mathcal T_1(\h)$. It is a Banach space under the trace norm. Hilbert-Schmidt and trace-class operators are compact operators.

Let $\{|e_n\rangle\}_{n\ge 1}$ be a fixed orthonormal basis of $\h$ and let $X\in\mathcal B(\h)$.  We define a new operator $T[X]\in\mathcal B(\h)$ by the relation
$$
\langle e_m| T[X]e_n\rangle = \langle e_n| Xe_m\rangle.
$$
The new operator $T[X]$ is called the {\em transpose of $X$}. Of course, $T[X]$ depends on the choice of the basis $\{|e_n\rangle\}_{n\ge 1}$. One can show that $\|T[X]\|_\infty=\|X\|_\infty$, so $T$ is a linear isometry on $\mathcal B(\h)$ and moreover, $T^2=\bbbone$.

An operator $X\in\mathcal B(\h)$ is said to be {\em positive}, written $X\ge 0$, if $\langle \psi |X \psi\rangle\ge 0$ for all $|\psi\rangle\in\h$. We have the following equivalence: $X\ge 0$ if and only if $X=X^\dag$ and the spectrum of $X$ satisfies ${\rm spec}(X)\subset [0,\infty)$. A {\em density matrix} is an operator $\rho\in \mathcal T_1(\h)$ such that $\rho\ge 0$  and $\tr\rho=1$.

Composite quantum systems are described by tensor products of Hilbert spaces. Let $\h_A$ and $\h_B$ be two separable Hilbert spaces and set 
$$
\h_{AB}=\h_A\otimes\h_B.
$$
We say that a density matrix $\rho$ on the bipartite Hilbert space $\h_{AB}$ is {\em separable} if it can be approximated in trace norm by a convex combination of product states \cite{Werner, CH}.\footnote{The word `separable' is used for states and, in a different context, for Hilbert spaces -- a separable Hilbert space is one which has a countable orthonormal basis. In the original paper \cite{Werner}, separable states are called {\em classically correlated} states.} That is, $\rho$ is separable if for $n\in\mathbb N$, there are density matrices $\rho_n^A$, $\rho_n^B$ on $\h_A$, $\h_B$, respectively, and probabilities $0\le p_n\le 1$, $\sum_{n\ge1}p_n=1$, such that 
\begin{equation}
	\label{a1}
	\rho = \sum_{n\ge 1} p_n \, \rho_n^A\otimes\rho_n^B,
\end{equation}
where the series converges in the $\|\cdot\|_1$ norm of $\h_{AB}$. If $\rho$ is not separable, then it is called {\em entangled}. Equivalently, the term inseparable is used \cite{HHH98}. An extraordinarily useful criterion to check that a state is entangled is the  PPT (positive partial transpose, or Peres-Horodecki) criterion, given in Theorem \ref{thm:PPT} below. Before stating it, we define the notion of partial transposition.

The {\em partial transposition} is the operation $\bbbone\otimes T$ acting on  $\mathcal B(\h_{AB})=\mathcal B(\h_A)\otimes\mathcal B (\h_B)$, where $T$ is the transposition operator on $\mathcal B(\h_B)$, relative to a fixed basis of $\h_B$, as introduced above. For $X\in\mathcal B(\h_{AB})$, the operator $(\bbbone\otimes T)X$ is called the {\em partial transpose} of $X$, also denoted by
$$
T_B[X] = (\bbbone\otimes T)X.
$$

A density matrix $\rho$ on $\h_A\otimes\h_B$ satisfying $T_B[\rho]\ge 0$ is said to be {\em positive partial transpose}, for short {\em PPT}. The following result is the famous PPT, or Peres-Horodecki criterion for separability, which originated in \cite{Peres, HHH96}.

\begin{thm}[PPT criterion]
	\label{thm:PPT}
	Let $\rho$ be a density matrix on the bipartite Hilbert space $\h=\h_A\otimes\h_B$. If $\rho$ is separable then $\rho$ is PPT.
\end{thm}

As discussed in the introduction, deriving the PPT property using perturbation theory is not immediate, in the infinite dimensional setting. However, for perturbations stemming from an interaction term in the Hamiltonian of an uncoupled bipartite equilibrium state, one can still use simple pertrubation theory to infer the PPT property. Let the Hamiltonian of a bipartite system, with Hilbert space $\h_A\otimes\h_B$, be given by 
\begin{eqnarray}
H_{AB} &=& H_0+V,\\
H_0 &=& H_A +H_B,
\end{eqnarray}
where $H_A\equiv H_A\otimes\bbbone_B$ and $H_B\equiv \bbbone\otimes H_B$ are individual (hermitian) Hamiltonians and $V$ is a hermitian interaction operator. It is assumed that for $\beta>0$,
\begin{equation}
\label{1.3}
    \tr_{AB}\, e^{-\beta H_0}<\infty,\quad  \tr_{AB}\, e^{-\beta H_{AB}}<\infty. \footnote{ If $\dim\h_{A,B}<\infty$, then \eqref{1.3} is automatically true. If $\dim \h_{AB} =\infty$,  then $\tr\, e^{-\beta H}<\infty$ (for $H=H_0$ or $H=H_{AB}$) is the same as saying that all the eigenvalues $\lambda_n > 0$ of $e^{-\beta H}$ are finitely degenerate and satisfy $\sum_n\lambda_n<\infty$; in particular, $\lambda_n\rightarrow 0$ as $n\rightarrow\infty$. Modulo a global additive shift, the energy spectrum of $H$ is given by $E_n=-\frac{1}{\beta}\ln(\lambda_n)$  and so $E_n\rightarrow\infty$, that is, $H$ must be unbounded.}
\end{equation}
The equilibrium state at inverse temperature $\beta$ is the Gibbs state with density matrix
\begin{equation}
\label{1.5-1}
\rho_\beta = \frac{e^{-\beta H_{AB}}}{\tr\, e^{-\beta H_{AB}}}\ .
\end{equation}

To carry out a rigorous proof, we make the following assumption:  There are constants $s_*>0$ and $a,b\ge 0$ such that for $0\le s\le s_*$,
\begin{equation}
\label{1.7}
\| e^{-s H_0}Ve^{sH_0} \|_2 \le a e^{bs}.
\end{equation}
Our main result is
\begin{thm}
\label{thm1}
Suppose $\beta$ is small enough, 
\begin{equation}
	\label{betastar}
0<\beta\le \max\{s_*,
\beta_*\},\qquad \mbox{where}\qquad \beta_*\equiv \frac{2}{b}\ln\big[1+\frac ba\frac{\ln 2}{2}\big].
\end{equation}
Then the equilibrium state 
$\rho_\beta$, \eqref{1.5-1}, is PPT. 
\end{thm}

The theorem says that if the temperature $1/\beta$ is large enough, then the coupled equilibrium state $\rho_\beta$ is PPT.

\bigskip

\noindent
{\bf Discussion of Theorem \ref{thm1}.} 
\medskip

\noindent
{\bf (D1) Weak coupling {\em vs.}~low temperature.}  In the case where the interaction operator carries an overall coupling constant $\lambda$, that is $H=H_0+\lambda V$, the constant $a$ in \eqref{1.7} is multiplied by $|\lambda|$. So the upper bound $\beta_*$ in Theorem \ref{thm1} is 
$$
\beta_*=\frac{2}{b}\ln\big[1+\frac{1}{|\lambda|}\frac{b}{a}\frac{\ln 2}{2}].
$$ 
For fixed $a,b$ we have $\beta_*\sim \ln|\lambda|^{-1}$ for small $\lambda$. This means that taking low enough temperatures ($\beta$ large enough) is equivalent to taking the coupling $\lambda$ small, for the condition of Theorem \ref{thm1} to hold. 
\medskip

\noindent
{\bf (D2) Examples for the validity of \eqref{1.7}.\ }
\noindent
\begin{itemize}
\item[(1)]  If $\dim\h_A$, $\dim\h_B<\infty$, then we can take $a=\|V\|_2$ and $b=\|H_0\|_\infty$. (Note, by a simple translation in energy which does not affect the equilibrium state, we can always assume that $H_0\ge 0$.)

%\color{cyan}
As a concrete example, we consider two interacting qubits $A$ and $B$, with 
$$
H_A=\tfrac12\omega_A\sigma_z, \ H_B=\tfrac12\omega_B\sigma_z, \ V=\lambda\sigma_x\otimes\sigma_x,
$$
where $\omega_A,\omega_B>0$ and  $\lambda\in\rx$ are constants and $\sigma_x, \sigma_z$ are Pauli matrices, with $\sigma_x=|-\rangle\langle +| \ +\  |+\rangle\langle -|$. Then 
$$
e^{-sH_0} V e^{sH_0} = \lambda \big(e^{-s\omega_A}|+\rangle\langle -| +  e^{s\omega_A} |-\rangle\langle +|\big)\otimes \big(e^{-s\omega_B}|+\rangle\langle -| +  e^{s\omega_B} |-\rangle\langle +|\big)
$$ 
and one calculates
$$
\|e^{-sH_0}Ve^{sH_0}\|_2 = 2\lambda \sqrt{\cosh(2s \omega_A)\cosh(2s\omega_B)}\le 2\lambda e^{s(\omega_A+\omega_B)}.
$$
Thus \eqref{1.7} is satisfied for all $s\ge 0$ with $a=2\lambda$ and $b=\omega_A+\omega_B$.
%\color{black}

\item[(2)] Denote the energy spectrum of $H_0$ by $E_j$, $j\ge 1$. We consider interaction operators $V$ having the following property: Starting from any $E_j$,  $V$ only makes transitions to finitely many levels $E_{j\pm \ell}$, where $\ell\in \{0,1,\ldots,L_j\}$ for some $L_j\ge 0$. Written in matrix form, $V$ is then a matrix whose nonzero entries on row $j$ lie within a neighbourhood of size $L_j$ around the diagonal entry,
\begin{equation}
V = \begin{pmatrix}
* & * & * &  & &  & & & & \\
 & * & * & * &* &  & & & &\\
 &  & * & * & & & & & & \\
 & * & * & * &* &  & & & &\\
  & & & & \ddots & & &  & &\\
  &  &  &  & *& * &* &   & &\\
 &  &  &  & &  *& *&* &*  &\\
 & & & &  & & &  \ddots & & \\
  &  &  &  & &  & &  & &
\end{pmatrix}
\end{equation}
In other words,
\begin{equation}
V = \sum_{j\ge 1} \sum_{\ell=1}^{L_j} V_{j,j+\ell}\,  |\phi_j\rangle\langle\phi_{j+\ell}| +{\rm h.c.},
\end{equation}
where $V_{j,k}=\langle\phi_j| V\phi_k\rangle$. Then one easily sees that 
$$
\| e^{-s H_0} Ve^{s H_0}\|_2 \le  2\sum_{j\ge 1}\sum_{\ell=1}^{L_j} e^{|s||E_j-E_{j+\ell}|} |V_{j,j+\ell}|.
$$
So we have the bound \eqref{1.7} for all $s\ge 0$, with 
\begin{equation}
\label{2.10}
a = 2\sum_{j\ge 1}\sum_{\ell=1}^{L_j}|V_{j,j+\ell}|,\qquad b=\sup_{j\ge 1}\max_{1\le \ell\le L_j} |E_j-E_{j+\ell}|.
\end{equation} 
For $a$ to be finite, we need $V_{j,j+\ell}\rightarrow 0$ with increasing $j$, meaning that the interaction has a high energy cutoff. This is a physically reasonable condition.

%\color{cyan}
As an explicit example, we consider the Jaynes-Cummings type Hamiltonian describing an atom (2 level system) interacting with a radiation mode (harmonic oscillator), described by  the Hamiltonians
\begin{equation}
\label{JC}
H_A = \omega \sigma_z,\  H_B = \Omega a^\dagger a,\ V= \chi_E (\sigma_+\otimes a + \sigma_-\otimes a^\dagger) \chi_E.
\end{equation}
Here, $\omega, \Omega >0$ and $\sigma_\pm$ are the raising and lowering operators of the atom. The $\chi_E$ is an energy cutoff, namely the projection operator onto the eigenspace of $H_0=H_A+H_B$ associated to eigenvalues $\le E$, where $E>0$ is any fixed number. Consider for simplicity of the presentation that $\omega <\tfrac12 \Omega$. Then the eigenvalues of $H_0$, denoted in increasing order as $E_1<E_2<E_3<\cdots$, are given by $E_{2n}= (n-1)\Omega+\omega$ and $E_{2n+1} = n\Omega-\omega$. The associated eigenvectors are $|\psi_{2n}\rangle=|+,n-1\rangle$ and $|\psi_{2n+1}\rangle =|-,n\rangle$, where $\pm$ in the ket refers to the ground and excited state of the atom and $n$ in the ket refers to the level of excitation of the oscillator. One readily checks that the interaction operator $V$ can only make transitions $E_k\rightarrow E_\ell$ with $|k-\ell|\le 2$. In other words, $L_j=2$ in the above language. As the interaction operator $V$ is essentially proportional to the square root of the oscillator number operator, we have $|V_{j,j+\ell}|\le 2\sqrt{j+\ell}$. Using this in \eqref{2.10} yields the bounds $a\le 8\sum_{j=1}^{J_E}\sqrt{j+2}$ and $b\le \Omega$, where $J_E$ is the largest integer satisfying $E_{J_E}\le E$. It is interesting to note that for the Jaynes-Cummings model \eqref{JC} {\em without} the cutoff $\chi_E$, the equilibrium state is not separable for {\em any} (nonzero) temperature, as was shown in \cite{Khemmani}.

%\color{black}

\item[(3)] A popular open system is an oscillator (system $a, a^\dag$) interacting with $N$ other oscillators (bath $b_j, b_j^\dag$) through the Hamiltonian 
\begin{equation}	
\label{H0}
H' =  H_0+ \sum_{j=1}^N \big( g_j a^\dag b_j +{\rm h.c.}\big),\qquad H_0 = \omega_0a^\dag a +\sum_{j=1}^N \omega_j b_j^\dag b_j,
\end{equation}
where $\omega_0,\omega_j >0$, the $g_j\in\mathbb C$ are coupling coefficients. A related model is the {\em spin Boson} model (with finitely many oscillators) and its generalizations, in which the system oscillator is replaced by a finite $L$-level system. Our method applies to that model as well. The interaction in \eqref{H0} is not a bounded operator so Theorem \ref{thm1} does not apply without modifications. 

One way to make the interaction bounded is to exclude processes involving energies beyond some cutoff $\Omega>0$ . This amounts to setting
\begin{equation}
	\label{V}
V = \chi_{H_0\le\Omega} \Big(\sum_{j=1}^N g_j a^\dag b_j +{\rm h.c.}\Big)\chi_{H_0\le\Omega},
\end{equation}
where $\chi_{H_0\le\Omega}$ is the spectral projection of $H_0$ onto the eigenspaces with spectral values contained in the interval $[0,\Omega]$.  We then consider the Hamiltonian
\begin{equation*}
H=H_0+ V,
\end{equation*}
where $H_0$ and $V$ are given in \eqref{H0} and \eqref{V}, respectively. We show in Section \ref{Ax} that 
\begin{equation}
	\label{1.12-2}
\| e^{-sH_0} Ve^{sH_0}\|_2 \le 2  \|g\|_2  \sqrt{N} (\Omega/\omega_{\rm min}+1)^{(N+3)/2}   e^{|s|\Delta\omega},
\end{equation}
where
$$
\omega_{\rm min}=\min_{0\le j\le N}\omega_j,\qquad \Delta\omega = \max_{1\le j\le N}|\omega_0-\omega_j|,\qquad \|g\|_2 = \Big( \sum_{j=1}^N|g_j|^2\Big)^{1/2}.
$$
The condition \eqref{1.7} holds with 
$$
a= 2  \|g\|_2 \sqrt{N} (\Omega/\omega_{\rm min}+1)^{(N+3)/2}  , \qquad b = \Delta\omega.
$$
The presence of low lying modes ($\omega_{\min}$ small) increases the value of $a$, and hence diminishes $\beta_*$, \eqref{betastar}. This means that if some of the oscillators have low frequencies, then the temperature $1/ \beta$ must be chosen large in order for Theorem \ref{thm1} to hold. 
\end{itemize}

\section{Proofs}

\subsection{Proof of Theorem \ref{thm1}}
\label{sect:proof}

Using the Dyson series we have
\begin{equation}
\label{1.4}
e^{-\beta H_{AB}} = \big[ \bbbone +D(\beta)\big] e^{-\beta H_0}
\end{equation}
with
\begin{equation}
\label{1.5}
D(\beta)= \sum_{n\ge 1} \int_0^\beta ds_1\int_0^{s_1}ds_2\cdots \int_0^{s_{n-1}}ds_n\  V(s_n)\cdots V(s_2)V(s_1)
\end{equation}
and where  
\begin{equation}
V(s) = e^{-s H_0} V e^{sH_0}.
\end{equation}
Under the condition \eqref{1.7}, the series \eqref{1.5} converges in the Hilbert-Schmidt norm and
\begin{eqnarray}
\|D(\beta)\|_2 &\le& \sum_{n\ge 1}a^n \int_0^\beta ds_1\int_0^{s_1}ds_2\cdots \int_0^{s_{n-1}}ds_n\  e^{b(s_1+\ldots+s_n)} \nonumber\\
&=& \sum_{n\ge 1}\frac{a^n}{n!} \big[\int_0^\beta e^{bs}ds\big]^n = \exp\big[a\frac{e^{\beta b}-1}{b}\big] -1.
\label{2.9}
\end{eqnarray}
We use \eqref{1.4} to arrive at
\begin{eqnarray}
e^{-\beta H_{AB}} &=&  \big[ e^{-\beta H_{AB}/2}\big]^\dag e^{-\beta H_{AB}/2} \nonumber\\
&=& e^{-\beta H_0/2}\big[\bbbone+D(\beta/2)^\dag\big] \big[\bbbone+D(\beta/2)\big] e^{-\beta H_0/2}\nonumber\\
&=& e^{-\beta H_0/2}\big[\bbbone +F(\beta) \big] e^{-\beta H_0/2},
\label{1.9}
\end{eqnarray}
where 
\begin{equation}
\label{2.6}
F(\beta) =   D(\beta/2)^\dag +D(\beta/2)+D(\beta/2)^\dag D(\beta/2).
\end{equation}
For any Hilbert-Schmidt operator $X$, we have 
$$
\|X^\dag\|_2=\|X\|_2\quad \mbox{and}\quad  \|X^\dag X\|_2\le \|X\|_2\|X\|_\infty\le \|X\|_2^2.
$$
Hence it follows from \eqref{2.9}, \eqref{2.6} that
\begin{equation}
\|F(\beta)\|_2 \le \|D(\beta/2)\|_2 \big(2+\|D(\beta/2)\|_2\big) \le \exp\big[2\, a\frac{e^{\beta b/2}-1}{b}\big] -1.
\label{1.11}
\end{equation}

Denote by $T_B$ the linear operator acting on operators of $\h_{AB}$, which takes the partial transpose of the system $B$. More precisely, let $|e_k\rangle$ and $|f_\ell\rangle$ be orthonormal bases of $\h_A$ and $\h_B$ and let $X$ be a bounded linear operator acting on $\h_{AB}=\h_A\otimes\h_B$. Then $T_B[X]$ is the linear operator on $\h_{AB}$, defined by its matrix elements
\begin{equation}
\langle e_k\otimes f_\ell | T_B[X] | e_m\otimes f_n\rangle = \langle e_k\otimes f_n |  X  | e_m\otimes f_\ell\rangle.
\end{equation}
The map $T_B$ depends on the choice of the basis of $\h_B$ -- which we consider to be arbitrary, but fixed. The map $T_B$ generally does not preserve the $\|\cdot\|_\infty$ norm of operators, but it leaves the Hilbert-Schmidt norm invariant,\footnote{To see that \eqref{2.14} holds, one notices that $T_B$ is (hermitian) selfadjoint with respect to the inner product of $\mathcal T_2(\h_{AB})$, namely for all $X,Y\in\mathcal T_2(\h_{AB})$, we have $\langle T_B[X]| Y\rangle=\langle X | T_B[Y]\rangle$. Then \eqref{2.14} follows by expressing the norm via the inner product and using that $T_B[T_B[X]]=X$.} 
\begin{equation}
\| T_B[X]\|_2 = \|X\|_2, \quad X\in\mathcal T_2(\h_{AB}).
\label{2.14}
\end{equation}
Moreover, we have 
\begin{equation}
\label{2.14-1}
(T_B[X])^\dag= T_B[X^\dag],
\end{equation}
so if $X$ is hermitian, then so is $T_B[X]$ and vice versa.

Let $X_A,Y_A$ and $X_B,Y_B$ be operators on $\h_A$ and $\h_B$, respectively, and let $Z$ be an operator on $\h_{AB}$. Then one readily sees that 
$$
T_B\big[(X_A\otimes X_B) Z (Y_A\otimes Y_B)\big] = (X_A\otimes T[Y_B])\,  T_B[Z]\,  (Y_A\otimes T[X_B]),
$$
where $T[\cdot]$ is the transpose in the given, fixed basis of $\h_B$. As $e^{-\beta H_0} = e^{-\beta H_A}\otimes e^{-\beta H_B}$, one obtains by applying $T_B$ to \eqref{1.9} that
\begin{equation}
T_B[e^{-\beta H_{AB}}] = \big\{e^{-\beta H_A/2}\otimes T[e^{-\beta H_B/2}]\big\}\,  T_B\big[\bbbone+F(\beta)\big] \, \big\{e^{-\beta H_A/2}\otimes T[e^{-\beta H_B/2}]\big\}.
\label{1.12}
\end{equation}
The operator $e^{-\beta H_A/2}\otimes T[e^{-\beta H_B/2}]$ is positive and invertible, and it follows from \ref{1.12} that\footnote{The implication $\Longleftarrow$ in \eqref{1.13} is immediate by \eqref{1.12}. In infinite dimensions, $\Longrightarrow$ needs to be shown with some care, since even though $Q\equiv e^{-\beta H_A}\otimes T[e^{-\beta H_B}]$ is a positive bounded operator, its inverse, $Q^{-1}\equiv e^{\beta H_A}\otimes T[e^{\beta H_B}]$ is an {\em unbounded} operator. Suppose then that
$Y\equiv T_B[e^{-\beta H_{AB}}]\ge 0$. We want to show that $X\equiv T_B[\bbbone+F(\beta)]\ge 0$. Let $P_n$ be the spectral projection of $Q$ on the subspace where $Q\ge 1/n$. Then $P_n YP_n=P_n QXQP_n$ and, as $Q^{-1}P_n$ is bounded, $Q^{-1}P_n YP_n Q^{-1}=P_nXP_n$. By the positivity of $Y$ and hence that of $Q^{-1}P_n YP_n Q^{-1}$, we have $\langle f|P_nXP_n|f\rangle \ge 0$ for any vector $|f\rangle$. Since $P_n|f\rangle\rightarrow |f\rangle$ as $n\rightarrow\infty$ it follows that $\langle f | X |f\rangle\ge 0$ for any vector $|f\rangle$. Hence $X\ge 0$.
} 
\begin{equation}
T_B[e^{-\beta H_{AB}}] \ge 0\quad\Longleftrightarrow \quad  T_B\big[\bbbone+F(\beta)\big] \ge 0.
\label{1.13}
\end{equation}
Next, $T_B[\bbbone+F(\beta)] = \bbbone+T_B[F(\beta)]$. As $T_B[F(\beta)]$ is hermitian, the spectrum of $\bbbone+T_B[F(\beta)]$ is real and we have the bound 
\begin{equation}
T_B[\bbbone+F(\beta)] \ge  \bbbone-\|T_B[F(\beta)]\|_\infty  \ge  \bbbone-\|T_B[F(\beta)]\|_2 =  \bbbone-\|F(\beta)\|_2. 
\end{equation}
Combining this with \eqref{1.11} we obtain
\begin{equation}
    T_B[\bbbone+F(\beta)]\ge 2-\exp\big[2a\frac{e^{\beta b/2}-1}{b}\big].
\end{equation}
The right hand side is $\ge 0$ provided 
\begin{equation}
\beta\le \frac{2}{b}\ln\big[1+\frac ba\frac{\ln 2}{2}\big].
\label{1.16}
\end{equation}
It follows that under the condition \eqref{1.16}, the operator $T_B[e^{-\beta H_{AB}}]$ is PPT. This completes the proof of Theorem \ref{thm1}. \hfill \qed

\subsection{Proof of \eqref{1.12-2}}
\label{Ax}

The eigenvectors of $H_0$  are $|\mathbf n\rangle\equiv |n_0,n_1,\ldots,n_N\rangle$, where  $\mathbf n = (n_0,n_1,\ldots,n_N)$ and $n_0, n_j\in\mathbb N$ are the occupation- or excitation numbers of the oscillators. The associated eigenvalues  are
$$
E(\mathbf n)= \omega_0 n_0 +\sum_{j=1}^N\omega_j n_j.
$$
We calculate
\begin{eqnarray}
	\| e^{-sH_0} Ve^{sH_0}\|_2^2 &=& \tr \big(e^{-sH_0}V e^{2sH_0} V e^{-sH_0}\big)\nonumber\\
	&=& \sum_{\mathbf n} e^{-2s E(\mathbf n)} \langle\mathbf n|Ve^{2sH_0}V|\mathbf n\rangle\nonumber\\
	&=& \sum_{\mathbf n, \mathbf m} e^{-2s [E(\mathbf n)-E(\mathbf m)]} |\langle\mathbf n|V|\mathbf m\rangle|^2.
	\label{1.12-1}
\end{eqnarray}
The operator $a^\dag b_j$ acts on an eigenstate as $a^\dag b_j |\mathbf m\rangle = \sqrt{(m_0+1)m_j} |\mathbf m'\rangle$, where $\mathbf m'$ is obtained form $\mathbf m$ by reducing $m_j$ by one and increasing $m_0$ by one. It follows that $\langle\mathbf n| a^\dag b_j|\mathbf m\rangle = \sqrt{(m_0+1)m_j} \delta_{n_0,m_0+1} \delta_{n_j,m_j-1}\prod_{\ell\neq 0,j} \delta_{n_\ell,m_\ell}$ (Kronecker deltas). Note also that $\chi_{H_0\le\Omega}|\mathbf m\rangle = \chi_{E(\mathbf m)\le\Omega}|\mathbf m\rangle$.  We then estimate the matrix element as
\begin{eqnarray}
	|\langle\mathbf n|V|\mathbf m\rangle|^2 &\le& \chi_{E(\mathbf m)\le\Omega} \ \chi_{E(\mathbf n)\le\Omega} \Big(\sum_{j=1}^N |g_j| \big\{  \langle \mathbf n| a^\dag b_j|\mathbf m\rangle  +  \langle \mathbf n| a b_j^\dag|\mathbf m\rangle \big\} \Big)^2\nonumber\\
	&\le& 2 \chi_{E(\mathbf m)\le\Omega} \ \chi_{E(\mathbf n)\le\Omega} \Big(\sum_{j=1}^N|g_j|^2\Big)
	\Big(\sum_{j=1}^N  \langle \mathbf n| a^\dag b_j|\mathbf m\rangle^2 +  \langle \mathbf n| a b_j^\dag|\mathbf m\rangle^2 \Big), \nonumber\\
	&
	\label{1.13-01}
\end{eqnarray}
where we used the Cauchy-Schwarz inequality for sums and that $(A+B)^2\le 2 (A^2+B^2)$. Next, 
\begin{eqnarray}
	\sum_{j=1}^N  \langle \mathbf n| a^\dag b_j|\mathbf m\rangle^2 +  \langle \mathbf n| a b_j^\dag|\mathbf m\rangle^2 &=& \sum_{j=1}^N (m_0+1)m_j \langle\mathbf n| \mathbf m'_j\rangle + m_0(m_j+1) \langle \mathbf n|\mathbf m_j''\rangle\nonumber\\
	&\le& \sum_{j=1}^N (m_0+1)(m_j+1) \big( \delta_{\mathbf n, \mathbf m'_j} +  \delta_{\mathbf n, \mathbf m_j''}\big)
	\label{1.14}
\end{eqnarray}
where $\mathbf m'_j$ is $\mathbf m$ with $m_0$ replaced by $m_0+1$ and $m_j$ replaced by $m_j-1$ and similarly for $\mathbf m''_j$. Here, $\delta_{\mathbf n,\mathbf k}$ takes the value $1$ if $\mathbf m=\mathbf k$ and $0$ otherwise. In view of \eqref{1.12-1} we need to multiply with $e^{-2s[E(\mathbf n)-E(\mathbf m)]}$. Either delta function selects values such that  $|E(\mathbf n)-E(\mathbf m)| =  |\omega_0-\omega_j|$; one excitation is transferred, and so  $e^{-2s[E(\mathbf n)-E(\mathbf m)]}\le  e^{2|s| |\omega_0-\omega_j|}$. Also,  the  summation over $\mathbf n$ in \eqref{1.12-1} disappears due to the presence of the delta functions.  We combine \eqref{1.12-1}-\eqref{1.14} to obtain
\begin{eqnarray}
	\label{1.15-1}
	\| e^{-sH_0} Ve^{sH_0}\|_2^2 &\le& 2
	\big(\sum_{j=1}^N|g_j|^2
	\big) \sum_{j=1}^N  e^{2|s| |\omega_0-\omega_j|} \nonumber \\
	&& \times \sum_{\mathbf m : E(\mathbf m)\le\Omega}(m_0+1)(m_j+1) \sum_{\mathbf n : E(\mathbf n)\le\Omega} ( \delta_{\mathbf n,\mathbf m'_j} +\delta_{\mathbf n,\mathbf m''_j})\nonumber\\
	&\le& 4\big(\sum_{j=1}^N|g_j|^2
	\big)  \sum_{j=1}^N  e^{2|s| |\omega_0-\omega_j|} \sum_{\mathbf m : E(\mathbf m)\le\Omega}(m_0+1)(m_j+1)\nonumber\\
	&\le & 4  \big( \sum_{j=1}^N|g_j|^2\big) N e^{2|s|\Delta\omega} (\Omega/\omega_{\rm min}+1) ^2 \sum_{\mathbf m : E(\mathbf m)\le\Omega}1,
 \label{3.19}
\end{eqnarray}
where $\omega_{\rm min} = \min_{0\le j\le N}\omega_j$ and $\Delta\omega=\max_{1\le j\le N}|\omega_0-\omega_j|$. We have used that in the summation, $\omega_k m_k\le\Omega$ for all $k=0,\ldots,N$. An easy (but rough) upper bound for the last sum in \eqref{1.15-1} is obtained %\color{cyan} 
as follows. That sum counts the number of indices $\mathbf m$ such that $E(\mathbf m)\le\Omega$ and we write it as
$$
\sum_{\mathbf m : E(\mathbf m)\le \Omega} 1 = \sum_{m_0\ge 0}\sum_{m_1\ge 0}\cdots \sum_{m_N\ge 0} \chi_{\sum_{j=0}^N \omega_j m_j\le \Omega}(\mathbf m),
$$
where $\chi_{\sum_{j=0}^N \omega_j m_j\le \Omega}(\mathbf m)$ is the function of $\mathbf m$ which equals one if the inequality is satisfied and zero else. In each one of the $N+1$ sums, we have $\omega_j m_j \le\Omega$, or $m_j\le \Omega/\omega_{\rm min}$. Let $M=\lfloor\Omega/\omega_{\rm min}\rfloor$ be the largest integer smaller than or equal to $\Omega/\omega_{\rm min}$. Hence
\begin{eqnarray*}
\sum_{\mathbf m : E(\mathbf m)\le \Omega} 1 &=& \sum_{m_0= 0}^M\cdots \sum_{m_N= 0}^M \chi_{\sum_{j=0}^N \omega_j m_j\le \Omega}(\mathbf m)\\
&\le& \sum_{m_0= 0}^M\cdots \sum_{m_N= 0}^M 1 = \Big(\sum_{m= 0}^M 1 \Big)^{N+1} = (M+1)^{N+1}\le (\Omega/\omega_{\rm min}+1)^{N+1}.
\end{eqnarray*}
Using this estimate in \eqref{3.19}
%\color{black}
we conclude that 
\begin{equation}
	\label{1.15-2}
	\| e^{-sH_0} Ve^{sH_0}\|_2 \le 2\sqrt{N} (\Omega/\omega_{\rm min}+1)^{(N+3)/2}  \Big( \sum_{j=1}^N|g_j|^2\Big)^{1/2} e^{|s|\Delta\omega}.
\end{equation}
This completes the proof of \eqref{1.12-2},\hfill\qed

\bigskip

{\bf Acknowledgements.} The work of both athours was supported by a Discovery Grant from NSERC, the National Sciences and Engineering Research Council of Canada. The authors are grateful to two anonymous referees for carefully reviewing this work and providing constructive feedback.
\medskip

{\bf Data availability.} 
Data sharing is not applicable to this article as no new data
were created or analyzed in this study.
\medskip

{\bf Competing interests.} 
The authors declare that there are no competing interests.

\end{document}